\documentclass[conference]{IEEEtran}
\IEEEoverridecommandlockouts

\usepackage{cite}
\usepackage{amsmath,amssymb,amsfonts}
\usepackage{algorithmic}
\usepackage{graphicx}
\usepackage{textcomp}
\usepackage{xcolor}
\def\BibTeX{{\rm B\kern-.05em{\sc i\kern-.025em b}\kern-.08em
    T\kern-.1667em\lower.7ex\hbox{E}\kern-.125emX}}

\usepackage{subcaption} 

\usepackage[most]{tcolorbox}

\newcounter{obscounter}
\renewcommand{\theobscounter}{\Roman{obscounter}}
\definecolor{custompurple}{HTML}{53257F}
\newtcolorbox{ObservationBox}[2][]{text width=0.94\linewidth,
colbacktitle=custompurple,enhanced,before upper={\parindent0mm\noindent},colback=custompurple!10!white,
attach boxed title to top left={yshift=-2mm,xshift=3mm},
boxed title style={sharp corners},top=6pt,bottom=2pt,
title=#2, left=4pt, right=2pt}
\newcommand{\obs}[1]{%
\refstepcounter{obscounter}%
\begin{ObservationBox}{\textbf{Observation \theobscounter}}%
#1%
\end{ObservationBox}}
    
\begin{document}

\title{TETRIS-Q: Tiling-based Effective Transient-fault Reduction on Interleaved Superconducting Qubits}

\makeatletter
\newcommand{\linebreakand}{%
  \end{@IEEEauthorhalign}
  \hfill\mbox{}\par
  \mbox{}\hfill\begin{@IEEEauthorhalign}
}
\makeatother

\author{
\IEEEauthorblockN{Marzio Vallero\IEEEauthorrefmark{1}}
\IEEEauthorblockA{marzio.vallero@unitn.it}
\and
\IEEEauthorblockN{Gioele Casagranda\IEEEauthorrefmark{1}\,\IEEEauthorrefmark{3}\,\IEEEauthorrefmark{4}}
\IEEEauthorblockA{gioele.casagranda@unitn.it}
\and
\IEEEauthorblockN{Flavio Vella\IEEEauthorrefmark{1}}
\IEEEauthorblockA{flavio.vella@unitn.it}
\and
\IEEEauthorblockN{Paolo Rech\IEEEauthorrefmark{2}\,\IEEEauthorrefmark{3}\,\IEEEauthorrefmark{4}}
\IEEEauthorblockA{paolo.rech@unitn.it}
\linebreakand
\IEEEauthorblockA{
\IEEEauthorrefmark{1}Department of Information Engineering and Computer Science, Univeristy of Trento, Italy\\
\IEEEauthorrefmark{2}Department of Industrial Engineering, Univeristy of Trento, Italy\\
\IEEEauthorrefmark{3}Department of Physics, University of Trento, Italy\\
\IEEEauthorrefmark{4}Trento Institute for Fundamental Physics and Applications, National Institute for Nuclear Physics, Italy
}
\thanks{
The authors acknowledge CINECA under the ISCRA initiative, for the availability of high-performance computing resources and support.
This work was partly supported by the INFN section 5 project \textit{QuRE}, by the Q@TN lab, and by the Italian Ministry for University and Research (MUR) through the Departments of Excellence 2023-27 program under Grant L.232/2016 awarded to the Department of Industrial Engineering.
}
}

\maketitle

\begin{abstract}
The struggle of the hour in quantum computing research is achieving effective suppression of the error mechanisms induced by the interaction of external radiation with superconducting quantum devices.
Despite the rapid advancements in quantum error correction (QEC) of recent years, radiation-induced faults are yet to be fully addressed.
These events are known to be the cause of simultaneous correlated defects in qubits that lie onto a single substrate, ultimately jeopardising QEC code effectiveness.

In this paper, we propose to \textit{selectively} combine substrate-level phonon barriers and QEC interleaving via a planar-mesh tiling algorithm, TETRIS-Q, reaching efficient and effective suppression of radiation events.
Our cross-layer solution comes at no extra cost in terms of QEC code execution or decoding time.
We model and simulate radiation-induced transient faults over a plethora of barrier and QEC interleaving configurations.
Through more than \textit{51 million} quantum circuit simulations, we show peak logical error reductions of more than $99.8 \%$, together with an $80\%$ reduction of the observable transient duration with permeable barriers.
We find that sparser tiling can reach comparable performance to \textit{single qubit} tiling, prompting cost reductions of upwards of $87 \%$ in barrier tracing.
By leveraging independent QEC code interleaving, we measure up to one order of magnitude average logical error rate reductions without the use of permeable barriers, and up to three orders of magnitude with the joint usage of barriers.
\end{abstract}


\section{Introduction}
\label{sec:intro}
Many promises have been leveraged over the past decades to support the development of quantum computers \cite{Shor1994, Grover1996, quantum-drug, Herman2023, Arute2019}.
Yet, at the present time, most of those promises are withheld from our grasp by notable engineering challenges; first and foremost, the achievement of fault tolerant quantum computers that suppress errors beyond the physical qubits' error rate.
Quantum Error Correction (QEC) strives to solve this very issue \cite{Bravyi2018, Bonilla_Ataides_2021, Acharya2023}, by employing a multitude of stabiliser and data qubits apt at dealing with device intrinsic noise, which generally manifests as a well defined set of non-correlated unitary transformations~\cite{Gidney2022intrinsicnoisemodel}.
Despite recorded successes in this \cite{Chen2021exponential,Acharya2023}, the lurking shadow of radiation-induced faults remains a widely undiscussed topic.
Few works have considered them, although the community has unequivocally identified radiation events to be one of the root causes of spatio-temporally correlated faults, which alone are sufficient to induce QEC failure \cite{vallero2024efficacy,Acharya2023,Acharya2024,Kurilovich2025correlated}.
Given that qubit's sensitivity to radiation is thousands times higher that that of CMOS transistors~\cite{Vepsalainen2020, Casagranda2025understanding}, it could potentially bind the large-scale adoption of quantum computers~\cite{dedominicis2024underground}.
Some of the solutions to radiation currently proposed in the literature either leverage underground facilities to reduce the external radiation flux \cite{dedominicis2024underground, Cardani2021, Cardani2023,Malevannaya2025shielding}, sever the superconductor-substrate coupling \cite{Junger2025suspendedqubits}, engineer higher energy gaps \cite{Acharya2024,Mannila2022traps}, or employ per-qubit quasiparticle barriers and traps \cite{Iaia2022, Martinis2021,Calusine2018trench,Murray2020trench,Henriques2019traps,Pan2022traps}.
However, shielding techniques are considerably impractical to scale, and can not prevent all radiation events from reaching the substrate, while gap engineering, suspended qubits, and quasiparticle traps incur in additional manufacturing costs and scalability constraints of superconducting quantum chips on a \textit{per qubit} basis.

In this paper, we aim at providing an effective and cost-efficient radiation-induced fault suppression by exploiting TETRIS-Q, a planar-mesh graph tiling algorithm, in order to combine modular barrier placement and QEC interleaving configurations.
With TETRIS-Q, we define a periodic and parametrisable method to subdivide the planar-mesh topology of quantum chip's qubits into separate and adjacent tiles, extensible over any quantum chip size.

By modelling the \textit{selective} usage and placement of substrate-level phonon barriers \cite{Iaia2022,Martinis2021} enclosing such tiles, we aim at investigating QEC's resilience to simulated radiation-induced faults.
Following real-world implementations, barriers are characterised by a permeability quality factor, which we take as input for our simulations, only partly limiting the dispersion of radiation-induced quasiparticles to well defined portions of the quantum chip.
The theory supporting phonon-barriers is that, by disrupting the spatially correlated nature of radiation faults, the resilience of current QEC codes will be considerably boosted, without incurring in any additional runtime or decoding overhead.

With ever increasing on-device qubit counts, more and more independent logical qubits are being embedded in independent QEC codes on a single chip.
We thus also leverage the TETRIS-Q tiling algorithm to interleave multiple separate QEC codes.
The intent is to increase the physical distance between virtually-close qubits in a QEC code whilst maintaining the same total number of embedded logical qubits. 

By merging phonon barriers placement and QEC interleaving, we manage to reduce the impact of simulated radiation-induced faults below the intrinsic noise floor of the superconducting quantum computer.
Besides, we optimise for the implementation cost of the phonon-barriers in order to reach the desired level of reliability without incurring in diminishing returns in effectiveness.

\noindent
In short, we ask ourselves:
\begin{itemize}
    \item \textbf{RQ1}: To what extent do phonon barriers improve QEC codes' tolerance to radiation events?
    \item \textbf{RQ2}: What is the optimal phonon barrier size and position to preserve QEC operativity?
    \item \textbf{RQ3}: Can phonon barriers improve the error threshold of QEC codes?
    \item \textbf{RQ4}: Does interleaving multiple independent QEC codes improve their reliability to radiation events?
    \item \textbf{RQ5}: What reliability improvements can be gained by using both barriers and QEC interleaving?
\end{itemize}

The manuscript is structured as follows.
In Section \ref{sec:background}, we provide a brief background on the effects of radiation events, other than introducing the QEC code object of our later analyses.
Following this, in Section \ref{sec:methodology}, we discuss the models used for the quantum computer, intrinsic noise and radiation-induced faults, then discuss the TETRIS-Q tiling strategy, the modelling of phonon barriers and the QEC interleaving approach.
We present our main findings in Section \ref{sec:results}, to then express our final considerations in Section \ref{sec:conclusions}.

\section{Background}
\label{sec:background}
In this section, we provide a description of both radiation events and QEC to acquaint the reader with the concepts of the later sections.


\subsection{Radiation events in quantum devices}
\label{subsec:radiation_events}
Being a well known source of faults in traditional electronics, numerous studies concerning reliability account for radiation events.
However, the scene for superconducting quantum computers is less populated, as the first studies addressing this same topic came to interest only very recently \cite{Vepsalainen2020,Wilen2021,casagranda2025squidgame}.
While a classical CMOS is generally susceptible only to bit-flip faults in the presence of considerable energy depositions from highly energetic particles \cite{Baumann2005}, superconducting qubits are much more sensitive.
This stems from the fact that the binding energy of the Cooper pairs flowing across these qubits is much smaller, and thus easily disrupted even by smaller energy depositions, a set of detrimental effects, also affecting other decoherence channels \cite{Thorbeck2023radiation}.
It has been experimentally observed that muons, which are generally considered harmless for classical electronics, end up being the first cause of radiation events \cite{Cardani2021, Casagranda2025understanding,Yelton2024}, given their larger abundance with respect to high energy neutrons at the surface of the Earth.
Similarly, photons can also be a source of correlated error bursts \cite{McJunkin2026ondemand}.
Radiation-induced events end up causing whole chip correlated errors with frequencies ranging from once every hour to multiple times per \textit{minute} \cite{mcewen2022resolving, Acharya2023,Yelton2024}.
The transient persistence of these effects has been recorded to last between microseconds and tens of seconds \cite{Oliveira2023neutrons}.
In the context of QEC, a round of correction lasts around a few microseconds or less, as such these events usually hinder multiple subsequent rounds of correction \cite{Acharya2023,vallero2024efficacy,Kurilovich2025correlated}.

\begin{figure}[!t]
    \centering
    \includegraphics[width=\linewidth]{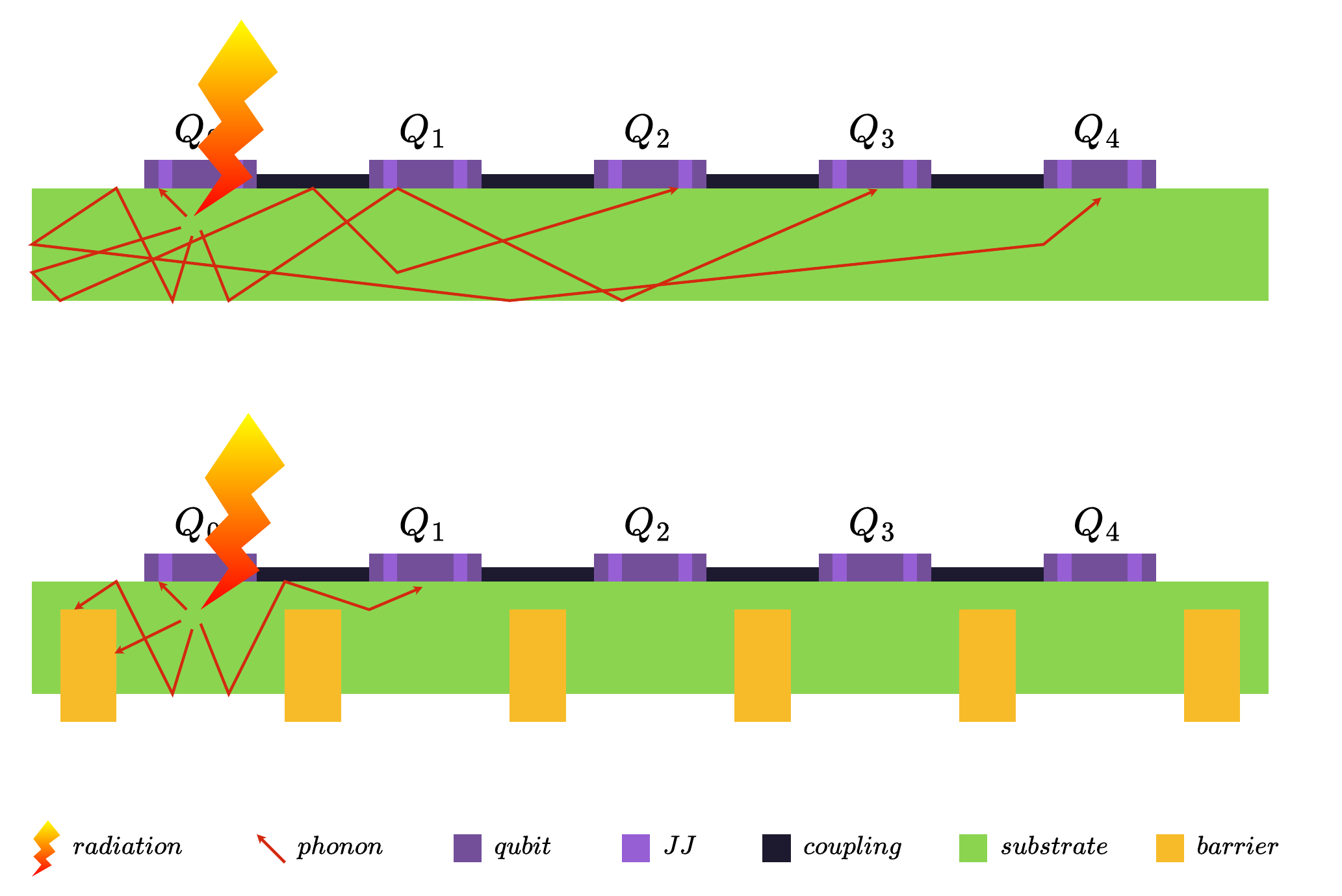}
    \caption{Simplified effect of substrate barriers, side view. Without barriers, quasiparticles can easily disperse throughout the quantum chip, causing correlated faults (top). Barriers confine radiation-induced quasiparticle dispersion (bottom).}
    \label{fig:teaser}
\end{figure}

\subsection{Quantum Error Correction (Algorithm-level)}
Quantum information, by definition, can not be cloned without prior knowledge of the encoded state \cite{Wootters1982nocloning}.
As such, QEC codes must be able to detect and correct errors through indirect measurements of the joint parity of multiple qubits.
To this end, a general QEC code is composed of a set of \textit{data} qubits, that hold redundant copies of the same information, and \textit{stabiliser} qubits, that perform parity measurements over well defined subsets of the data qubits over multiple rounds \cite{Chatterjee2023}.
This gives rise to a graph that relates the various stabiliser measurements in a code, which can later be leveraged to decode an \textit{error syndrome}.
The process of decoding involves making a prediction of the most likely generators of the measured syndrome, in order to issue corrective operators apt at reverting its effects and preserving the quantum information encoded therein \cite{Higgott2025sparse}.
QEC codes are thus developed considering non correlated sparse defects, while radiation events, by generating spatio-temporally correlated and locally dense defects, lie outside of the code space.
Recent research has explored the implementation of dynamic surface codes \cite{Eickbusch2025}, dynamic lattice surgery \cite{Leroux2025snakesladders} and super-stabilisers \cite{Wei2025superstabilizers} to improve the performance of QEC codes against isolated erasure errors. 
We are interested in overcoming these QEC limitations through cross-layer hardening employing substrate-level phonon barriers, which limit radiation faults' correlated and locally dense nature.
Moreover, we are also interleaving multiple QEC codes to increase the physical distance between virtually close qubits.

This study focused its efforts on the currently most widely adopted class of QEC codes, the rotated surface code \cite{Bonilla_Ataides_2021}.
The main advantage of this code class lies in its planar connectivity requirement amongst qubits of degree $\le 4$, which makes them easily implementable in current day quantum hardware.
This also gives rise to a defect decomposition that produces graph-like errors, which can be easily decoded with efficient and fast decoders, such as minimum weight perfect matching \cite{Higgott2022pymatching,Higgott2025sparse}.


\subsection{Radiation hardening methods (Hardware-level)}
Preliminary radiation-hardening solutions have been proposed.
Gap-engineering, an hardware-level tuning of the energy threshold required to cross the Josephson junctions, which is designed to regulate intrinsic device noise, has shown promising results in phonon-mediated cross talk reduction \cite{Wesdorp2026mitigating,Pinckney2026gapengineering,Binney2026distinguishing,Kamenov2024suppression}.
There is, however, limited evidence of its effectiveness over the whole spectrum of radiation events \cite{Kurilovich2025correlated,Acharya2024,Mannila2022traps,McEwen2024}.
Placing quantum computers in underground facilities limits the external radiation flux reaching the device, although scaling this approach is impractical \cite{Cardani2021,Cardani2023,dedominicis2024underground,Loer2024}.
The decoupling of superconducting components from the substrate has shown a notable reduction of the incidence of radiation events, but this also imposes serious manufacturing quality variance and scalability challenges \cite{Junger2025suspendedqubits}.
The usage of geometrical trenches, insulating layers, downconversion structures or phonon barriers limits the generation and diffusion of quasiparticles in superconducting quantum computers \cite{Calusine2018trench,Murray2020trench,Henriques2019traps,Pan2022traps,Iaia2022,Martinis2021,McEwen2024}, all while increasing manufacturing costs on a \textit{per qubit} basis, which makes them not cost-effective.
We thus propose to selectively engineer the utilisation of \textit{hardware-level} hardening by adapting it to \textit{algorithm-level} QEC.

\subsection{Main contribution}
The concept of substrate-level phonon barrier used in this paper refers to an hardware-level solution intent at limiting the spreading of radiation byproducts across the substrate of a quantum chip.
From an higher abstraction point of view, these barriers reduce the incidence of simultaneous correlated faults at the qubit level, as exemplified in Figure \ref{fig:teaser}.
We do not focus on any specific physical implementation for such barriers, as multiple alternatives are currently being explored in the field \cite{Iaia2022,Bargerbos2023mitigation}, but rather we look into the higher-abstraction implications of \textit{where} to put barriers, \textit{how many} qubits should they enclose, and \textit{how} to optimise their usage.
The aim is to provide both cost-efficient and effective radiation suppression via hardware-software co-design.

\obs{
Substrate barriers limit the propagation of radiation-generated phonons, aiming to reduce the incidence of spatially correlated radiation-induced faults.
} 

Any barrier is characterised by its ability to prevent the passage of quasiparticles, namely its \textit{permeability}.
The phonon flux attenuation provided by these barrier technologies has been experimentally measured to be in the order of $2-100\times$ \cite{Iaia2022,Bargerbos2023mitigation} with aluminium strip barriers with widths ranging between $10-100 \mu m$ \cite{Odeh2023nonmarkovian,Rosen2019protecting}, hinting at an inverse correlation between a barrier's width $w$ and its permeability, as $b_p \propto 1/w$.
In the context of a linear multiplicative model, a barrier with permeability $b_p = 1$ will not reduce the quasiparticle flux, while a barrier with permeability $b_p = 0$ is an ideal barrier, allowing no energy leakage.
Realistic engineering constraints limit the range of achievable barrier permeability quality factors as $b_p \in [0, 0.01]$, starting from more permeable trenches and metallisation stripes to less permeable multi-layered etchings.
We consider the increasing cost of implementing a less permeable barrier with respect to the increase in its width, in the following simple model correlating barrier width in $\mu m$ and the barrier permeability quality factor $b_p$ that reduces the phonon flux.
\begin{equation}
\label{eqn:barrier_permeability_cost}
w(b_p) = 10 \sqrt{\frac{1-x}{x}}
\end{equation}

The cost of implementing a barrier must then be multiplied by the total length of all the barriers that need to be manufactured to create a tiling of the whole quantum chip, as later explained in Section \ref{subsec:barriers}.

\begin{figure}[!t]
    \centering
    \includegraphics[width=\linewidth]{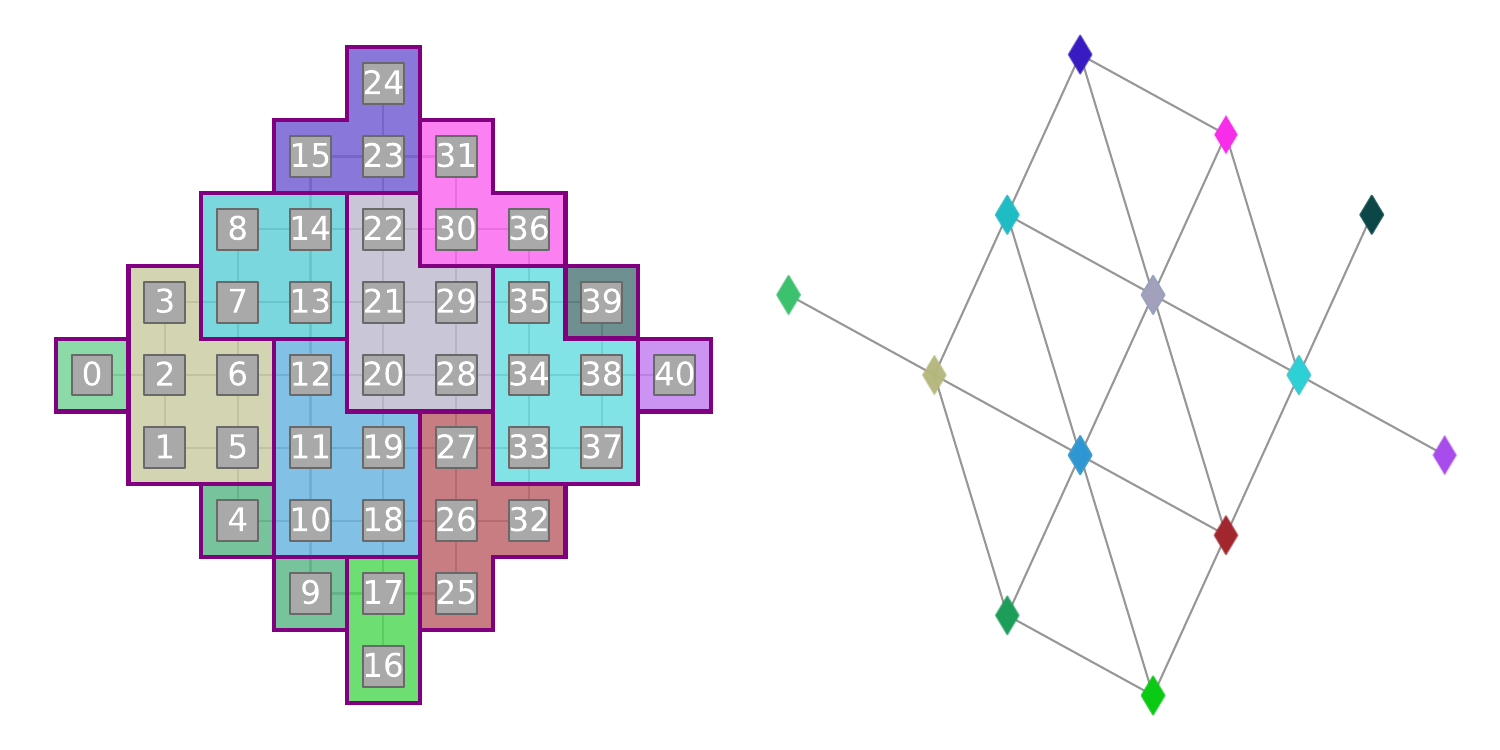}
    \caption{The quantum chip's topology (left) is used to create the barrier hypergraph (right) with a tiling pattern of dimension $tile_{size} = 5$, where each tile contains at most five qubits. The total phonon barrier perimeter is highlighted in purple.}
    \label{fig:tile_hypergraph}
\end{figure}

\begin{figure*}[!t]
    \centering
    \includegraphics[width=\linewidth]{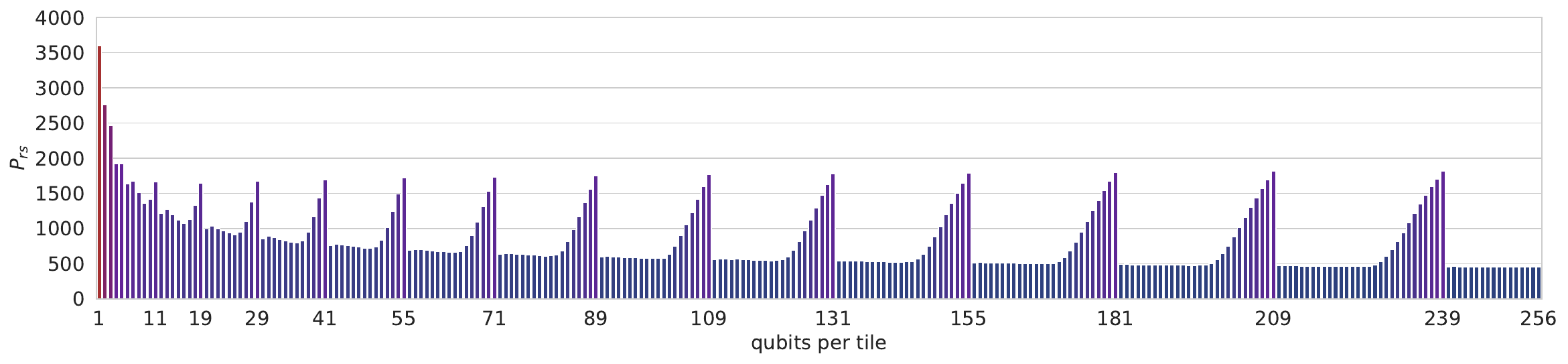}
    \caption{The $P_{rs}$ barrier cost per tile size, onto a square lattice topology boasting 1741 qubits. Due to quantum chip topology boundaries, the tiling cost does not decrease monotonically. X-axis labels represent local maximum $P_{rs}$ cost.}
    \label{fig:cost_per_tile_size}
\end{figure*}

We analyse the effectiveness of substrate-level phonon barriers by enclosing one or multiple qubits in tiles, an exclusive set of qubits which are spatially close on the quantum chip's substrate.
Quasiparticles can easily propagate inside a tile, while being less likely to diffuse across adjacent tiles.
In this context, a size-1 tile corresponds to a contiguous barrier enclosing a single qubit.
The subdivision of the qubits on the quantum chip in separate tiles of variable size is defined as a \textit{tiling} of the topology of the quantum chip.

\section{Setup and methodology}
\label{sec:methodology}
This section goes over intrinsic noise and radiation fault models used for our simulations.
We considered a generalised superconducting quantum computer model, with square lattice connectivity amongst a quantum chip with 1741 qubits.
This qubit count is not a stringent requirement, as most of the simulations required the usage of just a portion of the whole quantum chip.
All of the simulation data presented in this paper has been computed on the distributed nodes of the Leonardo supercomputer, provided by CINECA, Italy \cite{leonardo_2024}.
The simulation library employed is STIM \cite{Gidney2021stim}, while the radiation fault injector is to be made public as part of this paper's contributions.

\subsection{Intrinsic noise model}
For all our simulations, we considered the standard \textit{SI1000} intrinsic noise model, a commonplace representation of superconducting quantum computer inspired noise generally employed when testing QEC codes \cite{Gidney2022intrinsicnoisemodel}.
It operates by appending a randomised Pauli noise operator after each quantum gate in a quantum circuit, which is then triggered at the syndrome sampling step according to a given probability \textit{p}. 
The intensity of intrinsic noise is thus parameterised by \textit{p}, letting us easily sweep over a continuous range of intrinsic noise intensities.
With respect to the original implementation, the intrinsic noise model used in this paper has been extended to accommodate for all of the gates employed in the QEC codes we considered, but has been left otherwise unaltered.

\subsection{Radiation fault model}
Radiation events induce a spatio-temporally correlated reduction of the coherence time $\tau_1$ of multiple qubits \cite{Vepsalainen2020, mcewen2022resolving}.
The radiation fault model makes use of information from the properties of a modelled radiation event, the physical placement of qubits on the quantum chip, and the gate and coherence times of a generalised superconducting quantum computer \cite{Yelton2024,vallero2024efficacy,vallero2025detection,McJunkin2026ondemand}.
This model represents the effect of correlated radiation faults by injecting correlated Y-ERROR operators during the execution of a quantum circuit, which trigger according to a given probability.

The reduction in the characteristic $\tau_1$ time of a qubit follows $\tau_{rad}(t) = \tau_1 e^{10 \left( (t - t_{rad}) / \Delta t_{rad} - 1 \right)}$, taking into account the current time $t$, the beginning time of the radiation event $t_{rad}$, and the overall duration of the radiation event $\Delta t_{rad}$.
The $\tau_{rad}(t)$ coherence time is used as an argument of $T(\Delta t_g,t) = 1 - e^{- \Delta t_g / \tau_{rad}(t)}$, that together with the elapsed time since the last gate on a qubit, governs the probability of that qubit to undergo a radiation-induced fault.
The distance of a qubit $\Delta s$ from the impact point of the radiation event, also called locus of radiation, introduces the dampening factor $S(\Delta s) = 1 / (\Delta s + 1)^2$, that represents how energy dissipated in the quantum chip's substrate.
As such, the probability for a qubit to undergo a radiation-induced fault follows Equation \ref{eqn:rad_fault}.

\begin{equation}
\label{eqn:rad_fault}
P(\Delta t_g, t, \Delta s) = T(\Delta t_g,t) S(\Delta s)
\end{equation}

The radiation fault model automatically introduces these faults only during the $\Delta t_{rad}$ time window.

\section{Tiling, barriers and interleaving}
This Section describes the effects of barriers in the radiation-induced fault simulation model, and goes into detail over the tiling and interleaving strategy employed in this paper.

\begin{figure*}[!ht]
    \centering
    \includegraphics[width=0.95\linewidth]{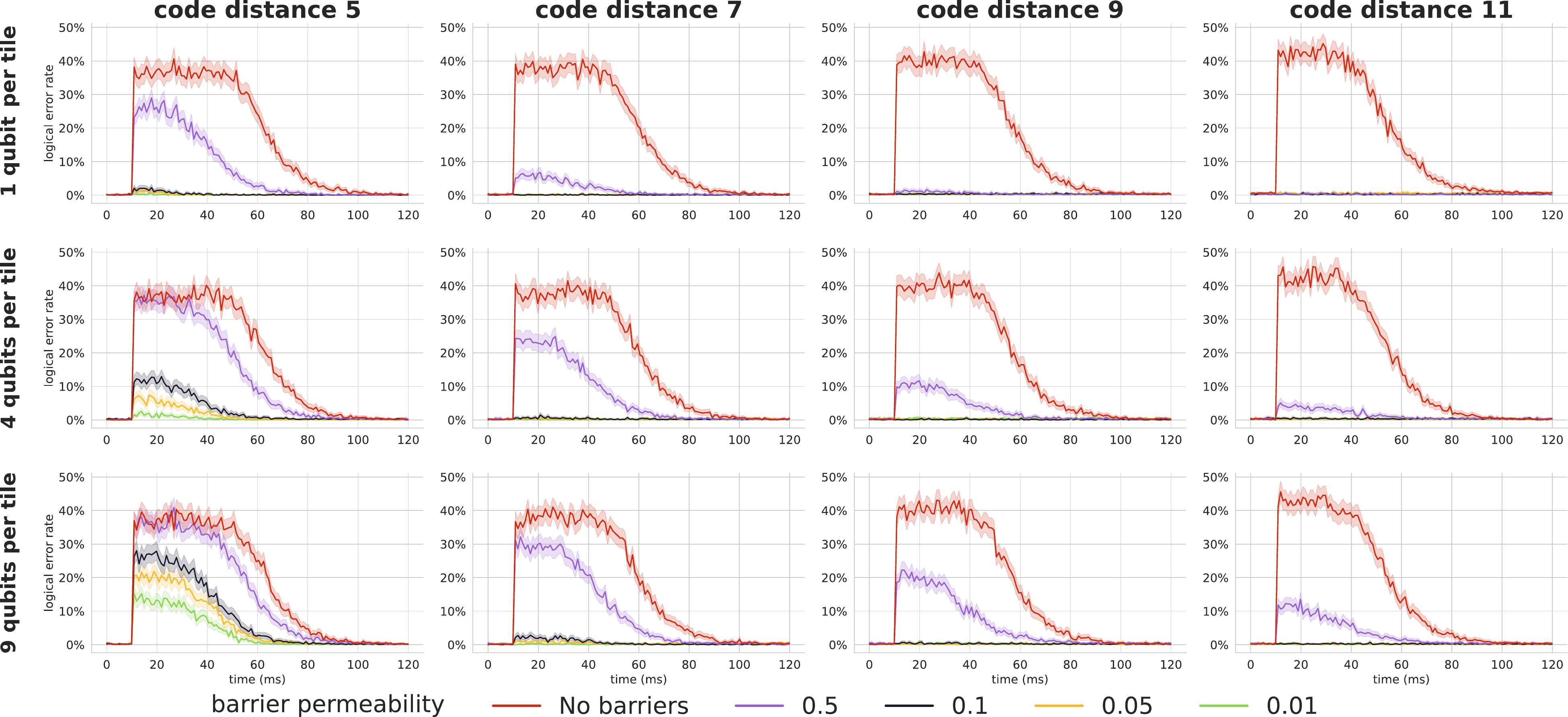}
    \caption{Rotated surface code logical error rates with 1, 4 and 9 qubit tiles (top to bottom), varying over increasing rotated surface code distance $d$ (left to right) and barrier permeability $b_p$ (plot hues). The case with no barriers acts as the base case, while barrier permeability costs follow Equation \ref{eqn:barrier_permeability_cost}. The tiling with 4-qubit tiles and 9-qubit tiles incur in barrier tracing costs of $47\%$ and $61\%$, respectively.}
    \label{fig:LER_over_time_by_class}  
\end{figure*}

\subsection{Tiling strategy}
The TETRIS-Q algorithm we used is a general algorithm working over a coordinate-aware planar square lattice graph.
The algorithm takes as input a graph, a starting vertex $v_s$, usually at the center of the graph, and a tile size $t_{size}$.
At first, it identifies the shape of the \textit{master tile}, which is composed of the largest complete square of vertexes, of side $s_{size} = \lfloor{\sqrt{t_{size}}\rfloor}$, plus the remaining vertexes $V_r = t_{size} - s^2_{size}$ wrapped around the upper-right side of the square.
The first tile is placed onto the coordinates associated to the vertex $v_s$, then all other non-overlapping tiles are discovered via a breadth-first search.
With this configuration, if multiple tiles are adjacent to one another, they share at least one side, and each tile contains the same number of vertexes, exception made for the vertexes on the outer portions of the graph.
The algorithm produces an hypergraph, where each vertex is a contraction of a group of vertexes from the input graph.

\subsection{Barriers}
\label{subsec:barriers}
Barriers are thus placed onto the perimeters of the tiles computed by the TETRIS-Q algorithm, which can easily extend the tiling to any quantum chip topology.
In the hypergraph, each vertex represents a tile, and the edges represent a shared barrier with a physically adjacent tile.
An example of tiling pattern and hypergraph generation is presented in Figure \ref{fig:tile_hypergraph}.
In the context of barriers $t_{size}$ directly amounts to the total number of qubits inside each tile.

In the context of barriers, each tile introduces a barrier on its perimeter.
This imposes one additional dampening factor on the probability of a qubit to experience a radiation-induced fault, as outlined in Equation \ref{eqn:barriers}, where $\Delta l$ is the path length between the tile of a qubit and locus of radiation's tile, and $b_p$ is the parameterisable barrier permeability.
Notably, a $b_p=1$ means that the barriers provide no damping.

\begin{equation}
\label{eqn:barriers}
B(\Delta l) = b_p^{\Delta l}, \quad\quad \Delta l \in \mathbb{N}
\end{equation}

This lets us extend Equation \ref{eqn:rad_fault}, adding the $B(\Delta t)$ damping parameter induced by the presence of tiles, giving rise to Equation \ref{eqn:rad_fault_barriers}, which has been used in this paper.

\begin{equation}
\label{eqn:rad_fault_barriers}
P(\Delta t_g, t, \Delta s, \Delta l) = T(\Delta t_g,t) S(\Delta s) B(\Delta l)
\end{equation}

The fabrication costs of barriers scale linearly with the total perimeter of all the barriers that need to be put onto the quantum chip.
To compute this, one must thus resort to the barrier hypergraph, since it provides information about the adjacency of barriers.
The total length $P_{rs}$ is thus given by the summation of the perimeter of all barriers, minus all the overlapping portions of the perimeter of each tile in excess of one.

\obs{
The cost of a tiling depends on the size and shape of the tile, and the relative placement of the tiling with respect to the quantum chip's topology.
}

In Figure \ref{fig:cost_per_tile_size} one can observe the cost $P_{rs}$, i.e. the total length of the barrier traces in relative units, in function of the tile size, on the quantum computer topology, for a total of 1741 physical qubits.
The maximum cost is obtained when each qubit has its separate tile, while the minimum cost is obtained for a tile size that encompasses all qubits.
Besides, the goal of tiling barriers is to simultaneously minimise both the $P_{rs}$ cost and the expected logical error rate of QEC, as we will discuss in Section \ref{sec:tile_size_permeability}.

\begin{figure*}
\begin{subfigure}[h]{0.5\linewidth}
    \centering
    \includegraphics[width=\linewidth]{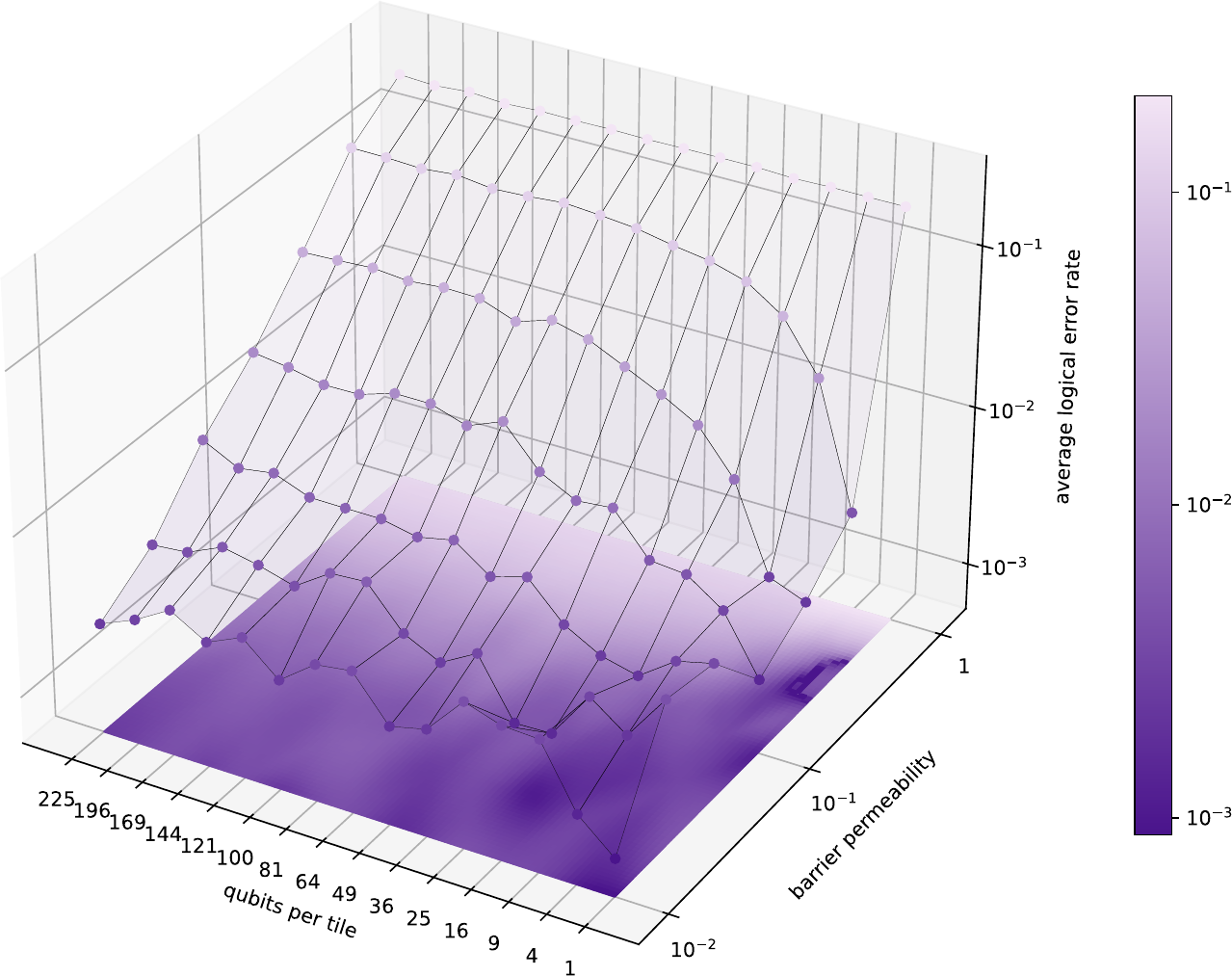}
    \caption{Non-centered phonon barrier tiling.}
\end{subfigure}
\hfill
\begin{subfigure}[h]{0.5\linewidth}
    \centering
    \includegraphics[width=\linewidth]{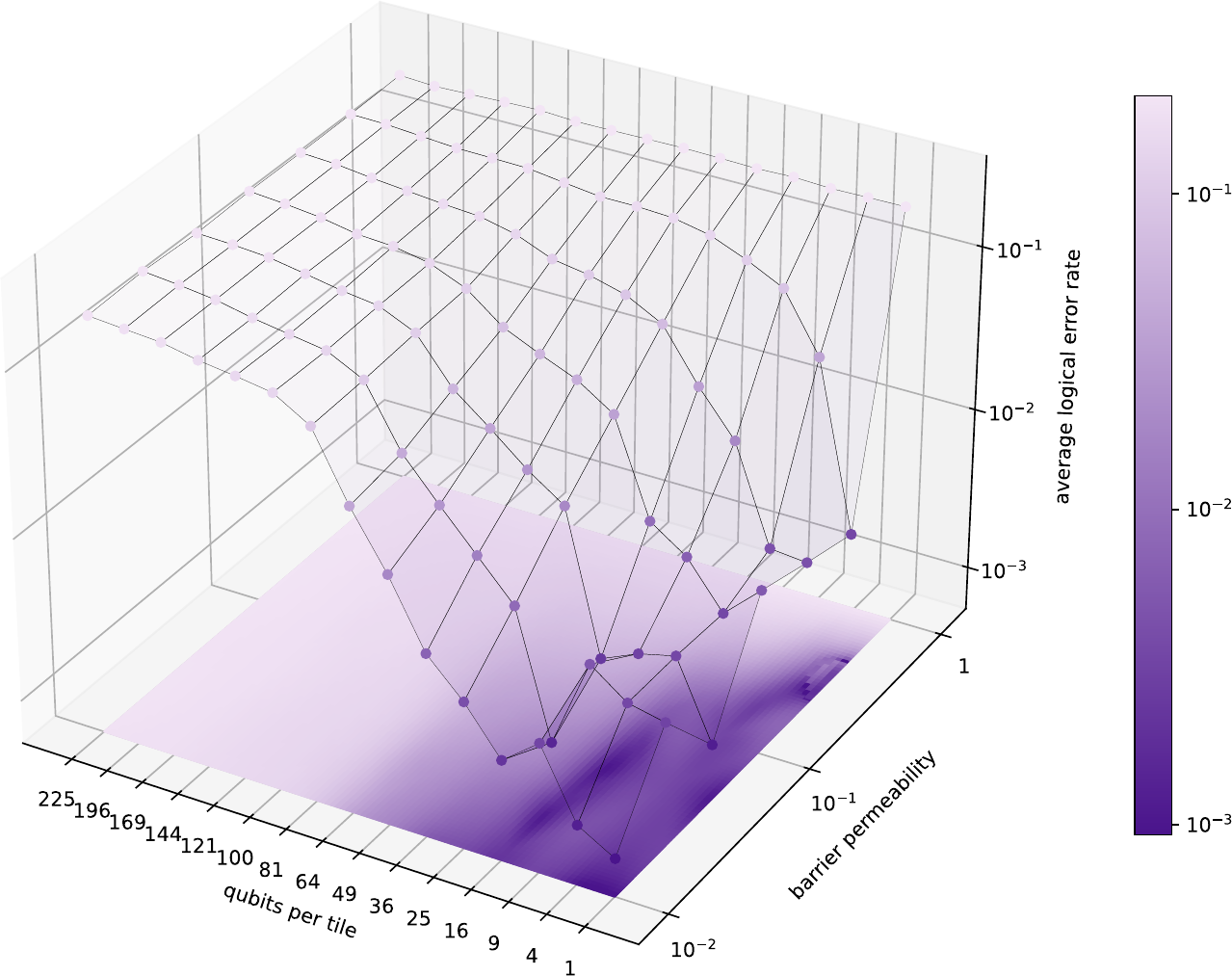}
    \caption{Centered phonon barrier tiling.}
\end{subfigure}%
\caption{Distance-9 rotated surface code average logical error rate (Z-axis, vertical), varying over qubits per tile (X-axis, left) and barrier permeability $b_p$ (Y-axis, right), differentiated by tiling positioning.}
\label{fig:average_LER_by_barrier_size}
\end{figure*}

\subsection{Interleaving}

Quantum circuit interleaving is a software-hardware co-design approach that aims at finding suitable quantum circuits mappings over a grid of qubits.
Given the rise of hardware-software co-design solutions for QEC codes, we see this as a further possible improvement to disentangle spatially close qubits from virtually correlated errors in QEC codes.
This specialisation is deemed extremely important in the context of future fault tolerant quantum computers, where multiple QEC codes, and thus logical qubits, may share the same purpose-built quantum chip.
From a topological standpoint, only qubits directly involved in the operation of a specific QEC code will be interconnected, thus giving rise to coupling line overlaps to be managed via three-dimensional coupling lines over the substrate \cite{Spring2022threedimensionalarchitecture,Saslow2025threedimensionalarchitecture,Dobrovolskiy2026Roadmap,AbuGhanem2025Roadmap}.
The implications of this choice from an engineering standpoint will not be made subject of analysis in this paper.
From an higher abstraction standpoint, interleaving independent QEC codes is the dual of the base case, which corresponds to relegate independent QEC codes in non-overlapping portions of the quantum chip.

The hypergraph can also be used to identify interleaving patterns to map multiple quantum circuits sharing the same dependency graph.
The spatial sorting of vertexes, i.e. qubits, in each tile of the hypergraph is preserved.
This makes it so that one can map a circuit onto the hypergraph derived from the quantum chip's topology, and then shift this mapping for multiple circuits among all of the positions inside each tile.
In a top-down view of the quantum chip, this amounts to interleaving multiple independent quantum circuits.

\section{Results}
\label{sec:results}
This Section goes over the main findings of the paper.
We considered a quantum computer with a characteristic $\tau_1$ time of $85$ $\mu s$, a single-qubit/two-qubit/measure-reset gate duration of $25/32/58$ $ns$.
The SI1000 \cite{Gidney2022intrinsicnoisemodel} intrinsic noise rate is set to $p=10^{-5}$, unless otherwise noted.
All the simulated radiation faults considered have been injected at the centre of the quantum chip, where the reach and intensity of the fault is maximal across the largest number of physical qubits.
The duration of the radiation fault is set to $100$ $ms$, following both results from the literature \cite{Wilen2021,Casagranda2025understanding,Cardani2023,casagranda2025squidgame,Yelton2024} and similar simulations of the same phenomena \cite{vallero2024efficacy,vallero2025detection}.
The QEC codes considered have been decoded using the PyMatching MWPM decoder \cite{Higgott2022pymatching}, as internal tests using other graph-based decoders showed no evident performance difference in the context of simulated radiation-induced faults.
The sampling resolution per time step has been uniformly set to $1024$ samples. In our observations, this was not a source of statistical variation.

\begin{figure*}[!ht]
    \centering
    \includegraphics[width=0.95\linewidth]{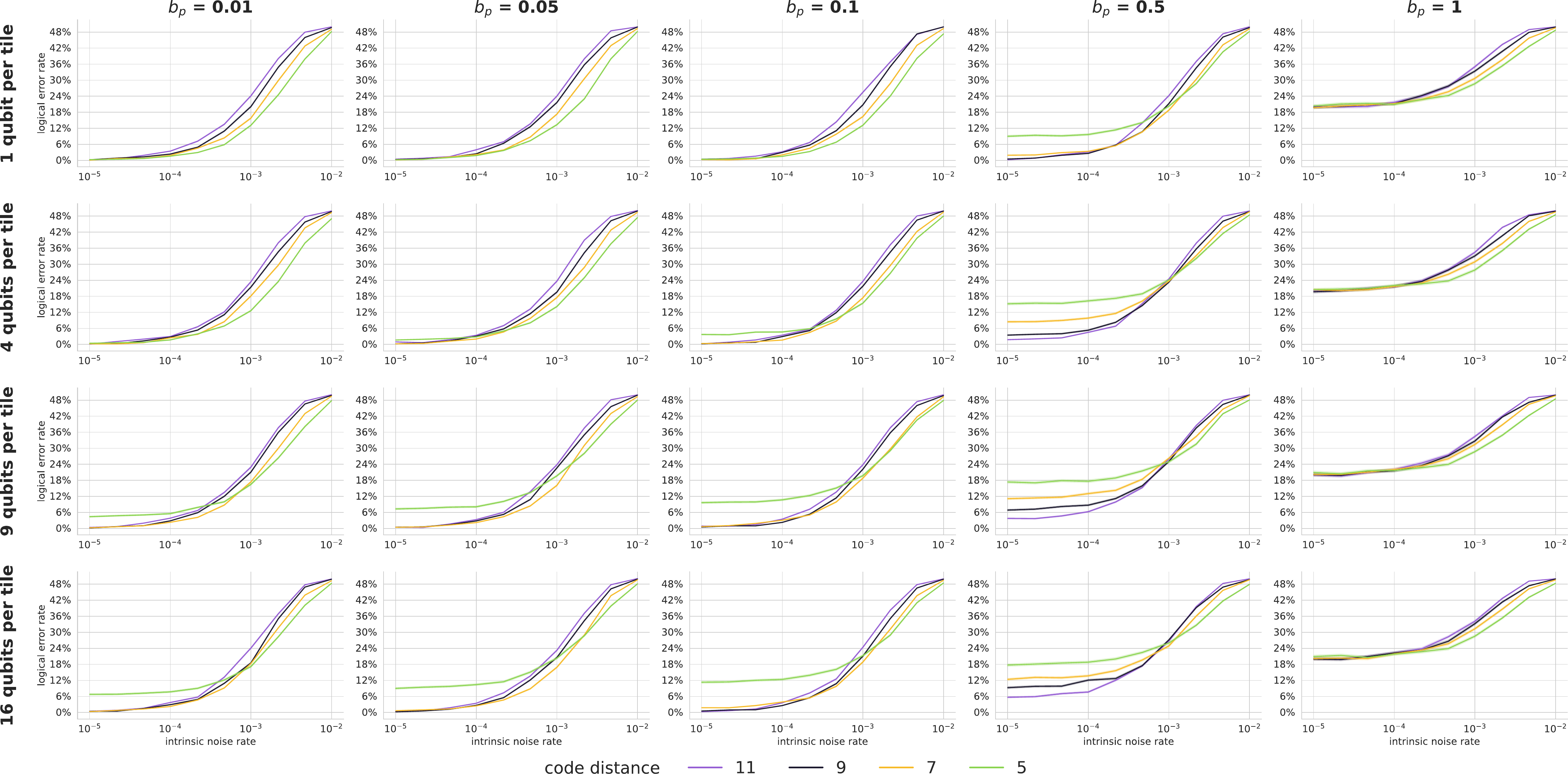}
    \caption{Average logical error rate of different rotated surface code distances with respect to intrinsic noise intensity, varying over barrier permeability $b_p$ (increasing from left to right) and $tile_{size}$ (increasing from top to bottom), averaged over the total duration of identical radiation-induced faults.}
    \label{fig:average_LER_over_noise_by_distance_and_permeability}
\end{figure*}

\subsection{Effect of barriers over time}

In this first analysis we consider a set of rotated surface code distances $d$, where $d \in \{5,7,9,11\}$, and measure the post decoding logical error rate in the presence of an identical radiation fault injected at $t=10$ $ms$ and lasting for $100$ $ms$.
The tiling patterns considered in this analysis are square shaped containing 1, 4 and 9 qubits, respectively, whilst performing a sweep over the barrier permeability factors with $b_p \in \{1, 0.5, 0.1, 0.05, 0.01 \}$.
We measure $141$ time steps and $1024$ shots per time step.

In Figure \ref{fig:LER_over_time_by_class}, we consider increasing phonon barrier tile sizes growing from top to bottom and rotated surface code distances growing from left to right, whilst logical error rate hues represent the barrier permeability used for that configuration.
In each configuration, we report the base case for barrier permeability $b_p=1$, equivalent to the base case with no barriers, characterised by a peak in the logical error rate at the beginning of the radiation event ($10$ $ms$) at about $40\%$.
This peak means that the QEC code is providing a performance slightly better than a random guess.
The logical error rate peak of this base case presents an inflection point at about $40 \%$ of the fault's duration, signalling a rapid reduction in the radiation fault's intensity, and a second inflection point at $65 \%$ of the fault's duration, with a more slowly fading tail until the end of the event.
With the introduction of single qubit barriers with a barrier permeability factor $b_p=0.1$, we notice a considerably smaller peak in the logical error rate of about $3\%$ for single qubit tiling at code distance-5, while the transient's duration dissipates just after $20\%$ of the temporal duration of the fault.
As the code distance increases, the effectiveness of barriers improves, with logical error rates not surpassing the $1\%$ threshold at distance-11, with a barrier permeability $b_p = 0.5$, whilst the same performance in the distance-5 code can only be reached with barrier permeabilities $b_p = 0.01$ more than one order of magnitude smaller.

\obs{
Phonon barriers produce observable effects even at relatively high permeability, reducing the peak and duration of the effects of radiation-induced faults.
}

In the case of a larger tiling pattern with 4 and 9 qubits per tile, we notice how smaller barrier permeabilities are required to keep the logical error rate under the $1\%$ threshold.
This is noticeable in the comparable performance of the distance-5 surface code on single-qubit tiles, the distance-7 code on 4-qubit tiles and the distance-9 code with 9-qubit tiles, all at the same barrier permeability $b_p = 0.5$.
This is due to the fact that by simultaneously increasing code distance and phonon barrier tile, each tile encloses a comparable percentage of all the qubits in the code.
With decreasing barrier permeability and phonon tile size and increasing code distance, the logical error rate peak and the overall witnessed duration of the radiation faults shrink under the code's threshold.
Notably, however, it is not necessary to do all those things at once to reach the desired radiation-induced fault response, prompting cost-saving strategies for what concerns phonon barrier tile sizes and permeabilities, and qubit requirements for code distances.
We performed other similar full timescale simulations for other tile sizes, which have not been reported for the sake of brevity, as the effect of tile size is discussed in the following Subsection.

\obs{
Lower barrier permeability and smaller phonon barrier tiles shunt the logical error rate peaks induced by radiation. 
}

\begin{figure*}[!ht]
    \centering
    \includegraphics[width=0.95\linewidth]{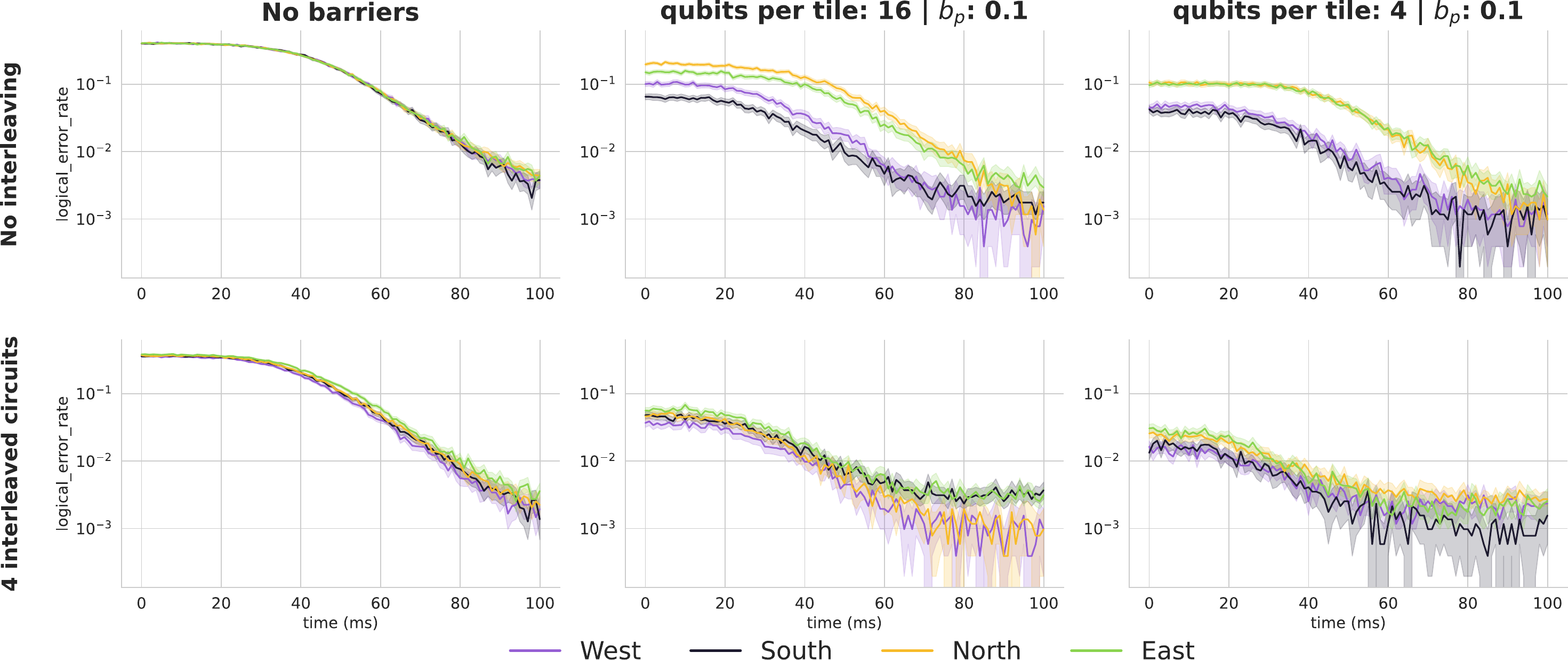}
    \caption{Logical error rate over time of four independent rotated surface codes subject to a radiation induced fault with respect the absence (top) or presence (bottom) of interleaving, varying over qubits per tile (from left to right), with a constant barrier permeability $b_p = 0.1$, where applicable.}
    \label{fig:LER_compare_barriers_and_interleaving}
\end{figure*}

\subsection{Effect of tiling size and position}
\label{sec:tile_size_permeability}

This second analysis investigates the effects of the size of square phonon barrier tiles sweeping over the perfect squares $\{ t_{size} \in \mathbb{N} : \sqrt{t_{size}} \in \mathbb{N}, t_{size} \le 225 \}$, in relation to the barrier permeability $b_p \in [1, 0.01]$ and the positioning of phonon barrier tiles.
We consider a distance-9 rotated surface code, and measure the average post-decoding logical error rate in the presence of an identical radiation fault.
We measure $20$ time steps and $1024$ shots per time step.

In Figure \ref{fig:average_LER_by_barrier_size}, we consider the case of growing square along two edges only (a), and the case of growing square tiles from all edges simultaneously (b), in both cases by starting the tiling from the center of the quantum chip.
In both subfigures, we sweep over the X-axis (left) with the number of qubits per tile, and over the Y-axis over the barrier permeability $b_p$ (right), whilst representing the average logical error rate during a radiation-induced fault on the Z-axis.
The barrier permeability has the most noticeable impact on the average logical error rate, with a stark correlation in both the non-center (a) and center (b) growing phonon barrier tiles.
This is confirmed by the fact that, when the barrier permeability $b_p=1$, the average logical error rate remains constant, regardless of the barrier size.
The placement of phonon barrier tiles is especially important when larger tiles are taken into consideration.
When the square tiles are centered with respect to the rotated surface code (b), the size of the phonon barrier tile show an improvement with respect to the baseline only if a single tile encloses at most $50\%$ the surface code, otherwise no gain is recorded, regardless of the barrier permeability.
In fact, the distance-9 rotated surface code employs 188 physical qubits, and for QEC-centered phonon barrier tiles holding more than 100 qubits we observe no benefit.

\obs{
Regardless of barrier permeability, centered phonon barrier tiles must enclose less than $50\%$ of the QEC code's qubits to produce observable effects.
}

Non-centered phonon barrier tiling is considerably more effective, as with the growing size of the phonon barrier tiles, each tile holds at most about $25\%$ of all the qubits in the rotated surface code.
Intuitively, the lowest average logic error rate of $10^{-3}$ is reached with single qubit tiles and the lowest barrier permeability $b_p = 0.01$ in both cases.
To reach an average logical error rate lower than $10^{-2}$ with a barrier permeability $b_p = 0.1$, it is sufficient to employ non-centered tiles containing at most 25 qubits, with a $P_{rs}$ cost reduction of more than $75\%$ with respect to tiles of size one.
Centered tiles reach the same average logical error rate only with 4-qubit tiles at the same barrier permeability.
This difference becomes even more evident with barrier permeability $b_p = 0.01$, where by using non-centered tiling, an average logical error rate approaching $10^{-3}$ can be guaranteed with tiles of up to 225 qubits, prompting a cost reduction of more than $87\%$.
As such, phonon barrier tile placement has a fundamental impact in the effectiveness of radiation-induced fault tolerance.
Through further similar simulations, not reported in this manuscript for the sake of brevity, we concluded that the maximal effectiveness of a barrier is achieved for a tile sizes containing less than a quarter of the QEC code's total qubits.

\obs{
Phonon barrier tiles containing up to $25 \%$ of a QEC code's qubits are comparatively effective to single-qubit phonon barrier tiles.
}

\subsection{Effect of barrier permeability with respect to noise}

In this third analysis, we correlate intrinsic noise and the average logical error rates over the whole duration of a radiation-induced fault.
Specifically, we sweep over the intrinsic noise model's probability $p\in[10^{-5}, 10^{-2}]$, for codes of distance $d \in \{5, 7, 9, 11 \}$, and phonon barrier tiles sizes $tile_{size} \in \{ 1, 4, 9, 16 \}$, to identify the rotated surface code's threshold when affected by radiation.
We measure $20$ time steps per configuration, with $1024$ samples per time step.

\begin{figure*}[!ht]
    \centering
    \includegraphics[width=0.95\linewidth]{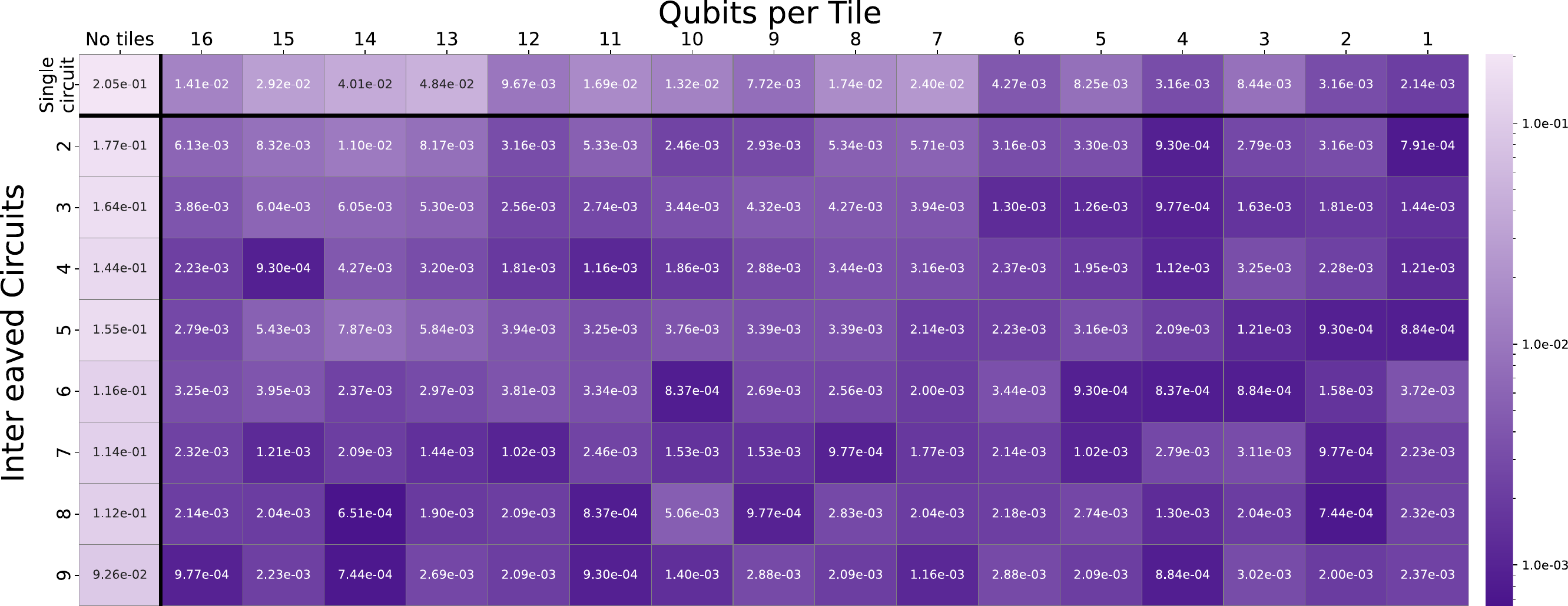}
    \caption{Average logical error rate of a distance-7 rotated surface code with respect to the number of interleaved independent surface codes (rows) and the number of qubits enclosed in each phonon barrier (columns) with a constant barrier permeability $b_p = 0.1$, averaged over the total duration of identical radiation-induced faults.}
    \label{fig:average_LER_barrier_interleaving_Rotated_surface_X_MWPM}
\end{figure*}

In Figure \ref{fig:average_LER_over_noise_by_distance_and_permeability} we identify the code thresholds in the considered combinations of barrier permeability and phonon barrier tile size.
If a QEC's codes logical error rate threshold under the effects of radiation persists beyond the intrinsic noise rate, the QEC code becomes non operational, accumulating more errors than can be corrected, as stated by observations \cite{vallero2024efficacy}.
In the base case with barrier permeability $b_p = 1$, corresponding to fully permeable barriers, the crossing points amongst the considered code distances is at $2.1 \times 10^{-1}$ over an intrinsic noise intensity of $10^{-4}$.
By introducing barriers with permeability $b_p = 0.5$, the crossing point stabilises between $1.8 \times 10^{-1}$ and $2.4 \times 10^{-1}$ with a one-order of magnitude more intense intrinsic noise rate.
Further reducing the barrier permeability correlates to lowering the code's threshold under the effect of radiation towards an error threshold comparable to the intrinsic noise intensity.

\obs{
Phonon barrier tiling can preserve the characteristic efficiency threshold of QEC codes to intrinsic noise, guaranteeing radiation-fault tolerance.
}

Given the temporally bound nature of radiation-induced faults, QEC codes can bear a temporary increase in the logical error rate past the nominal error threshold.
Although the best performing combination of barrier permeability and phonon barrier tiling size corresponds to single-qubit tiles with $b_p = 0.01$, the combination with the same barrier permeability and 4-qubits per tile performs comparatively well, with a $48\%$ reduction in the implementation cost of the tiling.
The distance-5 code responds less effectively to phonon barrier tiles, given its lower number of stabiliser qubits, and thus its crossing point with other code distances happens at a larger intrinsic noise rate.
When considering only code distances $d \in \{ 7, 9, 11 \}$, the crossing point reaches much better regimes, comparable with the single-qubit tiles with $b_p = 0.01$, with 9-qubits per phonon barrier tile and a barrier permeability one order of magnitude larger, prompting a cost saving of more than $60 \%$ and guaranteeing operativity under the effects of radiation.

\obs{
A QEC code's distance limits lower-permeability phonon barriers' efficacy, regardless of the intrinsic noise intensity.
}

\subsection{Effect of barriers and interleaving}
In the fourth analysis, we move towards the concept of quantum circuit interleaving, observing the logical error rate over the duration of a radiation-induced fault of four independent distance-7 rotated surface codes.
Specifically, we compare the base case without any interleaving, with the four interleaved QEC codes, over three different scenarios: a quantum chip without phonon barriers, a quantum chip with square tiles containing 16 qubits and a barrier permeability of $0.1$, and a quantum chip with square tiles containing 4 qubits and the same barrier permeability.
For each scenario, we take $1024$ samples per time step, and $100$ time step samples.

In Figure \ref{fig:LER_compare_barriers_and_interleaving}, the top row are the base cases with the four rotated surface codes placed side by side, whilst the bottom row are the cases with the four interleaved quantum circuits.
When comparing the two cases with no barriers, by using interleaving, the logical error rate peak sees a reduction of about $3\%$, but most importantly, the inflection point that signals the start of the dissipation of the radiation-induced fault appears $7\%$ earlier in the total duration of the fault.
This effect becomes even more evident with the introduction of phonon barriers.

\obs{
Phonon barrier tiles and QEC interleaving show positive interference in mitigating radiation-induced faults.
}

With a tiling containing 16 qubits, the peak logical error rate gets reduced by more than $10\%$ with the usage of interleaving, with an anticipation of the inflection point of up to $40\%$ of the total duration of the radiation-induced fault.
The third case, with a smaller tiling of 4 qubits, anticipates the position of the second logical error rate inflection point, after which the logical error rate reaches the logical error rate floor mandated by the intrinsic noise rate, by about $20\%$ with respect to the 16-qubits tiling and of about $40\%$ with respect to the 4 qubit tiling without interleaving. 

\subsection{Optimising barrier and interleaving cost}
This last analysis measures the average logical error rate of one distance-7 rotated surface code over a sweep from $1$ (base case) to $9$ interleaved quantum circuits and $b_p = 0.1$ phonon barriers with tile sizes ranging from $1$ to $16$ qubits, plus the base case with no barriers.
For each combination, we consider $1024$ samples per time step, and $20$ time step samples.

In Figure \ref{fig:average_LER_barrier_interleaving_Rotated_surface_X_MWPM}, the base case without phonon barriers presents the worst performance, acting as the comparison baseline.
Following the columns of the topmost row from left to right indicates ever smaller tiling for the phonon barriers, and likewise a reduction in the average logical error rate of the rotated surface code, up to a minimum of $2.14 \times 10^{-3}$ in the rightmost case of single-qubit tiles.
Nonetheless, the efficacy of tiling does not scale linearly with the number of qubits contained therein, as the tiles which present a square or rectangular shape are more effective at preventing the spread of phonons.
This can be noticed in the cases where the number of qubits per tile is either a perfect square, or the summation of the square of a perfect square with itself.

\obs{
Minimising the number of adjacent phonon barrier tiles improves their effectiveness.
}

The first column presents the isolated effect of interleaving without phonon barriers, reaching up to an $11\%$ reduction of the average logical error rate when $9$ surface codes are interleaved with respect to the base case.
The rest of the heatmap presents all of the combinations of tiling dimensions and number of interleaved QEC codes.
Notably, the best combinations of interleaving and phonon barrier tiles do not necessarily require a large number of interleaved QEC codes and small tiling patterns.
In the context of the quantum chip we modelled, the best average logical error rate of $6.51 \times 10^{-4}$ is achieved for 14-qubits phonon barrier tiles with 8 interleaved rotated surface codes.
However, by interleaving 4 rotated surface codes, and using phonon barriers around 15 qubits tiles, we reach an average logical error rate of $9.3 \times 10^{-4}$ during a radiation-induced fault, a performance which is comparable to the combination of single-qubit tiles and two interleaved rotated surface codes, whilst boasting a reduction of more than $70\%$ of the total cost of implementing phonon barriers on-chip.

\obs{
The cost of phonon barrier tiles can be further reduced via interleaving and the usage of larger distance codes, whilst reaching comparable radiation-induced fault tolerance.
}

Radiation event tolerance can thus be reached by employing a combination of both quantum circuit interleaving and phonon barrier tiling without incurring in prohibitively costed solutions.

\section{Conclusions and future works}
\label{sec:conclusions}
We have modelled and simulated substrate barriers in a superconducting quantum chip, with the intent of reducing the spatial correlation of radiation-induced faults in QEC codes.
Through a tiling of the device's substrate, group one or multiple physical qubits in tiles delimited by a barrier.
This barrier reduces the rate at which energy deposited by radiation events can spread across the quantum chip, without incurring in any additional overhead for QEC codes.
We considered the rotated surface code, measuring variations in the logical error rate across multiple code distances, barrier permeability factors, and intrinsic noise intensities.

Linking back to the research questions posed in Section \ref{sec:intro}, we characterised the immediate effect of barriers in limiting the temporal duration of radiation-induced faults, over a range of barrier permeability factors (RQ1).
As the barrier permeability diminishes, so does the peak logical error rate, eventually reaching the QEC code's characteristic noise floor (RQ1).
We highlighted how tiling size, up to one quarter of the QEC code's qubits, is sufficient to reach appreciable logical error rate reduction, prompting cost reductions, while tile positioning has an even more fundamental barrier efficacy impact (RQ2).
Moreover, we have correlated the benefits of lowering the barrier permeability and lowering the intrinsic noise rate of the quantum computer, showing that a QEC code's distance impacts the efficacy of barriers, but can not improve their intrinsic noise error threshold (RQ3).
As such, less permeable barriers should be employed together with less intrinsically noisy quantum hardware to observe significant improvements (RQ3).
We also considered the co-design approach of interleaving of multiple independent QEC codes, with the aim of increasing the spatial separation of qubits that refer to the same QEC code.
This method proved to be a viable to improve radiation-induced fault tolerance, reducing both the peak logical error rate and the persistence of the fault's temporal tail (RQ4).
The usage of both phonon barrier tiles and interleaving shows positive interference, further limiting the effects of radiation-induced faults below the considered rotated surface codes' thresholds (RQ5).

This work underlines the importance of addressing radiation-induced faults, providing a model with simulation results supporting a solution fit for addressing their correlated nature.
Albeit the results hereby presented are very promising, other approaches must also be investigated.
This includes other phonon barrier tiling methods, hardware hardening approaches, radiation-aware decoding-techniques and QEC codes, lattice surgery, and super-stabilisers methods, which are to be made object of future analyses. Together, these may come to play a role in ultimately extirpating the scourge of radiation-induced faults from superconducting quantum computers. 

\clearpage
\bibliographystyle{IEEEtran}
\bibliography{IEEEabrv,bibliography}

\end{document}